\documentclass[graybox]{svmult}

\usepackage{helvet}         % selects Helvetica as sans-serif font
\usepackage{courier}        % selects Courier as typewriter font
\usepackage{makeidx}         % allows index generation
\usepackage{graphicx}        % standard LaTeX graphics tool
\usepackage{multicol}        % used for the two-column index
\usepackage[bottom]{footmisc}% places footnotes at page bottom

\usepackage{amsmath, amssymb, amsfonts}

\def\cG{{\mathcal{G}}}

\def\hV{{\widehat V}}

\def\Z{\mathbb{Z}}
\def\C{\mathbb{C}}
\def\R{\mathbb{R}}

\def\H{\mathbb{H}}

\def\PSL{\mathrm{PSL}}

\def\dd{\mathrm{d}}

\def\vol{\mathrm{vol}}

\newcommand{\be}{\begin{equation*}}
\newcommand{\ee}{\end{equation*}}
\newcommand{\ben}{\begin{equation}}
\newcommand{\een}{\end{equation}}
\newcommand{\beqa}{\begin{eqnarray*}}
	\newcommand{\eeqa}{\end{eqnarray*}}
\newcommand{\beqan}{\begin{eqnarray}}
\newcommand{\eeqan}{\end{eqnarray}}

\newcommand{\Tr}{\mathrm{Tr}}
\newcommand{\tr}{\mathrm{tr}}

\def\tphi{\widetilde{\varphi}}

\def\odd{\mathrm{odd}}

\def\hSigma{\widehat{\Sigma}}

\def\G_2{\mathrm{G_2}}

\def\cL{\mathcal{L}}

\def\fI{\mathfrak{I}}
\def\fD{\mathfrak{D}}

\def\P{\mathbb{P}}

\def\G{\mathrm{G}}

\def\Im{\mathrm{Im}}

\def\grad{\mathrm{grad}}

\def\rS{\mathrm{S}}

\def\Re{\mathrm{Re}}
\def\Im{\mathrm{Im}}

\def\i{\mathbf{i}}

\def\tV{{\widetilde  V}}
\def\tv{{\widetilde  v}}
\def\tvarphi{{\widetilde \varphi}}

\def\cK{\mathcal{K}}

\newcommand{\eqdef}{\stackrel{{\rm def.}}{=}}

\begin{document}

\title*{Two-field cosmological models and the uniformization theorem}
\author{Elena Mirela Babalic and Calin Iuliu Lazaroiu}
\institute{Elena Mirela Babalic \at Center for Geometry and Physics, Institute for
  Basic Science, Pohang 37673, Republic of Korea, \email{mirela@ibs.re.kr}
\and Calin Iuliu Lazaroiu \at Center for Geometry and Physics, Institute for
  Basic Science, Pohang 37673, Republic of Korea, \email{calin@ibs.re.kr}}
\maketitle

\abstract{We propose a class of two-field cosmological models derived
  from gravity coupled to non-linear sigma models whose target space
  is a non-compact and geometrically-finite hyperbolic surface, which
  provide a wide generalization of so-called $\alpha$-attractor models
  and can be studied using uniformization theory.  We illustrate
  cosmological dynamics in such models for the case of the hyperbolic
  triply-punctured sphere.}

\section{Introduction}
\label{sec:1}

Inflation in the early universe can be described reasonably well
by so-called cosmological $\alpha$-attractor models \cite{Escher, Linde, Linde3}, 
which provide a good fit to current observational results.
The observational predictions of these models are to a large
extent determined by the geometry of the scalar manifold
rather than by the scalar potential.

The best studied $\alpha$-attractor models contain a single scalar
field, being obtained by radial truncation of two-field models based
on the hyperbolic disk \cite{Escher}. The latter arise from
cosmological solutions of 4-dimensional gravity coupled to a non-linear
sigma model whose scalar manifold is the open unit disk endowed with
its unique complete metric $\cG$ of constant negative Gaussian
curvature $K(\cG)=-\frac{1}{3\alpha}$, where $\alpha$ is a positive
parameter. As shown in \cite{Escher}, the ``universal'' behavior
of such models in the radial one-field truncation close to the
conformal boundary of the disk is a consequence of the hyperbolic
character of $\cG$, when the scalar potential is ``well-behaved'' near
the conformal boundary. It is thus natural to consider {\em
  two-field} $\alpha$-attractor models in which the hyperbolic disk is
replaced by an arbitrary hyperbolic surface $\Sigma$ which is
geometrically finite in the sense that its fundamental group is
finitely-generated.

\begin{definition} {\rm \cite{alpha}}
A {\em generalized two-field $\alpha$-attractor model} is defined by a
triplet $(\Sigma,\cG, V)$, where $(\Sigma,\cG)$ is a {complete}
geometrically-finite hyperbolic surface and $
V:\Sigma\rightarrow \R$ is a smooth potential function, while
$K(\cG)=-\frac{1}{3\alpha}$ with $\alpha>0$.
\end{definition}

\noindent This class of models is very rich. Since in general $\Sigma$
has non-trivial topology, a complete understanding requires going
beyond one field truncations. Instead, one can use the theoretical and
numerical methods of \cite{PT1,Mul} and certain other
approximation techniques \cite{alpha}.

\section{Cosmological models with two real scalar fields minimally coupled to gravity}
\label{sec:2}

Let us recall the general description of cosmological models with two
real scalar fields minimally coupled to gravity, allowing for scalar
manifolds of non-trivial topology, in a global and coordinate-free
description.

\subsection{Einstein-Scalar theories with 2-dimensional scalar manifolds} 

Let $(\Sigma,\cG)$ be any oriented, connected, complete
and possibly non-compact two-dimensional Riemannian manifold without
boundary called the {\em scalar manifold} and
$V\!:\!\Sigma\!\rightarrow \!\R$ be a smooth function called the {\em
  scalar potential}. We require completeness of the metric $\cG$ in
order to avoid problems with conservation of energy. For applications
to cosmology, it is important to allow $(\Sigma,G)$ to be non-compact
and of possibly infinite area.

Any triplet $(\Sigma,\cG, V)$ as above allows one to define an
Einstein-Scalar theory on any four-dimensional oriented manifold $X$
which admits Lorentzian metrics. This theory includes 4-dimensional
gravity (described by a Lorentzian metric $g$ defined on $X$) and a
smooth map $\varphi:X\rightarrow \Sigma$ (which locally describes two
real scalar fields), with action:
\ben
\label{S}
S[g,\varphi]=\int_X \cL(g,\varphi) \vol_g~~,
\een
where $\vol_g$ is the volume form of $(X,g)$ and $\cL(g,\varphi)$ is
the Lagrange density:
\ben
\label{cL}
\cL(g,\varphi)= \frac{M^2}{2} \mathrm{R}(g)-\frac{1}{2}\Tr_g \varphi^\ast(\cG)- V\circ \varphi~~.
\een
Here $\mathrm{R}(g)$ is the scalar curvature of $g$ and $M$ is the
reduced Planck mass. The quantity $\varphi^\ast(\cG)$ is the
pull-back through $\varphi$ of the metric $\cG$, while
$\Tr_g\varphi^\ast(\cG)$ denotes the trace of the tensor field of type
$(1,1)$ obtained by raising one of the indices of $\varphi^\ast(\cG)$
using the metric $g$.  The coordinate-free formulation \eqref{cL} allows one to
define such a theory globally for any topology of $\Sigma$ and $X$. 
The  last two terms in the  Lagrange density \eqref{cL}
describe the non-linear sigma model with source $(X,g)$,
target space $(\Sigma,\cG)$ and scalar potential $ V$.

\subsection{Cosmological models defined by $(\Sigma,\cG, V)$}
\label{subsec:FLRW}

By definition, a {\em cosmological model} defined by
$(\Sigma,\cG, V)$ is a solution of the equations of motion of the
theory \eqref{S}-\eqref{cL} when $(X,g)$ is a FLRW universe and
$\varphi$ depends only on the cosmological time. We assume for
simplicity that the spatial section is flat and simply connected. 
Hence the cosmological models of interest are defined by
the following conditions:
\begin{enumerate}
\itemsep 0.0em
\item $X$ is diffeomorphic with $\R^4$, with global coordinates
  $(t,x^1,x^2,x^3)$.
\item The squared line element of $g$ has the form:
\be
\label{FLRW}
\dd s^2_g=-\dd t^2+a(t)^2\sum_{i=1}^3{(\dd x^i)^2}~~,~~\rm{with}~~a(t)> 0~.
\ee
\item $\varphi$ depends only on $t$.
\item $(a(t),\varphi(t))$ are such that $(g,\varphi)$ is a solution
of the equations of motion
 derived from \eqref{S}.
\end{enumerate}
With these assumptions, the cosmological equations of motion are:
\be
 \label{eom}
\nabla_t \dot{\varphi}+3H \dot{\varphi}+(\grad_{\cG}  V)\circ \varphi =0~, ~~
\frac{1}{3}\dot{H}+H^2 - \frac{ V\circ \varphi}{3M^2} = 0~,~~
\dot{H} +\frac{\dot{\sigma}^2}{2M^2} = 0~,
\ee
where $\,\dot{~}\eqdef \frac{\dd}{\dd t}~,~\nabla_t\eqdef
\nabla_{\dot{\varphi}(t)}$ is the covariant derivative with respect to
$\dot{\varphi}(t)$, $\sigma$ is the proper length parameter on the curve
$\varphi(t)$, while $H\eqdef \frac{\dot{a}}{a}$ denotes the Hubble
parameter. The {\em inflationary regions} of a trajectory
$\varphi(t)$ are defined as the time intervals for which the scale
factor $a(t)$ is a convex and increasing function of $t$ ($\dot a>0,
~\ddot a>0$) and are given by the condition: \be
H(t)<H_c(\varphi(t))~, \ee where $H_c(p)\eqdef
\frac{1}{M}\sqrt{\frac{V(p)}{2}}$ is the {\em critical Hubble
  parameter} at a point $p\in \Sigma$. Approximations useful
for studying such models are discussed in \cite{alpha}.

\subsection{Two-field generalized $\alpha$-attractor models} 

\noindent By definition, a {\em hyperbolic surface} is a connected,
oriented, borderless and complete Riemannian two-manifold $(\Sigma,G)$
of constant Gaussian curvature equal to $-1$. A {\em two-field
  generalized $\alpha$-attractor model} is a two-field cosmological
model defined by a triple $(\Sigma, \cG, V)$ as above, where
$\cG=3\alpha G$ with 
$\alpha$ a positive parameter and $(\Sigma, G)$ is a hyperbolic surface.

\section{Uniformization of hyperbolic surfaces}

An isometric model of the Poincar\'e disk is provided by
the Poincar\'e half-plane, defined as the upper half-plane
$\H\eqdef \{\tau\in \C|~\Im \tau>0\}$ endowed with its unique
complete hyperbolic metric $ \dd s_\H^2=\lambda^2_\H(\tau,\bar{\tau})\dd
\tau^2$, where $\lambda_\H(\tau,\bar{\tau})=\frac{1}{\Im\tau}$. The
orientation-preserving isometries of $\H$ form the projective
special linear group $\PSL(2,\R)$. An element $A\in \PSL(2,\R)$ is
called {\em elliptic} if $|\tr(A)|<2$. By definition, a {\em surface
group} is a discrete subgroup $\Gamma$ of $\PSL(2,\R)$ which contains
no elliptic elements. Our analysis of generalized $\alpha$-attractor
models is based on the uniformization theorem \cite{unif}:

\begin{theorem}
For any hyperbolic surface $(\Sigma,G)$ there is a surface group
$\Gamma$ and a holomorphic covering map (uniformization map)
$\pi_\H:\H\longrightarrow\Sigma$ defined on the Poincar\'e half-plane
$\H$ such that $\Sigma$ is isometric to the quotient $\H/\Gamma$.
\end{theorem}

\noindent In this theorem, holomorphicity of $\pi_\H$ is understood
with respect to the complex structure $J$ induced on $\Sigma$ by the
conformal class of $G$. 

\subsection{Lifted trajectories and tilings} 

Consider the generalized $\alpha$-attractor model defined by a
hyperbolic surface $(\Sigma, G)$ at a fixed positive value of the
parameter $\alpha$. To study the cosmological trajectories
$\varphi:\fI\longrightarrow \Sigma$ (where $\fI\subset \R$ is an
interval), it is convenient to first study their lifts
$\tvarphi:\fI\longrightarrow\H$ to the hyperbolic plane and then
project them back to $\Sigma$ as $\varphi=\pi_\H\circ\tvarphi$.
The projection from $\H$ to $\Sigma$ can be determined if we know the
tiling of $\H$ determined by a fundamental polygon of $\Gamma$.  There
is no fully general stopping algorithm known for computing fundamental
polygons of surface groups. However, a general algorithm \cite{polyg} is known when
$\Gamma$ is an arithmetic surface group such that $\H/\Gamma$ has
finite hyperbolic area. The connection to uniformization theory shows
that the study of generalized $\alpha$-attractor models requires
sophisticated results from uniformization theory. When $\Sigma$ has
finite hyperbolic area, this is closely connected to the theory of
modular forms and hence to number theory.

\subsection{The end and conformal compactifications}

A non-compact hyperbolic surface $(\Sigma, G)$ has two natural
compactifications, namely the {\em end compactification} \cite{ends} of
Freudenthal and Kerekjarto-Stoilow (which depends only on the topology
of $\Sigma$) and the {\em conformal compactification} (which depends
on the conformal class of the hyperbolic metric $G$).  In the
geometrically-finite case, these two compactifications can be
described as follows:
\begin{itemize} 
\itemsep 0.2em
\item Since $\pi_1(\Sigma)$ is finitely-generated, $\Sigma$ is
  homeomorphic with $\hSigma \setminus \{p_1,\ldots, p_n\}$, where
  $\hSigma$ is a closed oriented surface and $p_1,\ldots,p_n$ are
  distinct points on $\hSigma$. The compact surface $\hSigma$ can be
  identified with the {\em end compactification} of $\Sigma$, while
  the points $p_1,\ldots, p_n$ can be identified with the {\em ends}
  of $\Sigma$.
\item As shown by Maskit, the {\em conformal compactification}
  $\bar\Sigma$ of $\Sigma$ (with respect to the complex structure $J$)
  can be identified with the topological closure of $\Sigma$ inside a
  closed Riemann surface which is obtained from $\Sigma$ by adding a
  finite number $n_c$ of points and a finite number $n_f$ of disks,
  where $n_f+n_c=n$. The topological boundary $\partial_\infty
  \Sigma=\bar\Sigma\setminus \Sigma$ consists of $n_c$ isolated points
  and $n_f$ disjoint simple closed curves and is called the {\em
    conformal boundary} of $\Sigma$. Contracting each of the $n_f$
  curves to a point recovers the end compactification, the $n_c$
  isolated points and the $n_f$ contraction points recovering the ends
  of $\Sigma$. Accordingly, the ends of $\Sigma$ divide into $n_c$
  {\em cusp ends} (those corresponding to points in the conformal
  compactification) and $n_f$ {\em flaring ends} (those corresponding
  to simple closed curves in the conformal compactification).
\end{itemize}

\noindent On the neighborhoods of each end, the hyperbolic metric can
be brought to one of four explicitly known forms (namely for the ``cusp'',
``funnel'', ``horn'' or ``plane'' end), thus providing the {\em isometric
  classification of ends}. 

\subsection{Well-behaved scalar potentials}

Let $\hSigma$ be the end compactification of $\Sigma$.  A scalar
potential $ V:\Sigma\rightarrow\R$ is called {\em well-behaved} at an
end $p\in\hSigma\setminus\Sigma$ if there exists a smooth function
$\hV_p:\Sigma\sqcup\{p\}\rightarrow\R$ such that $ V=\hV_p|_\Sigma$~.
The potential $ V$ is called {\em globally well-behaved} if it is
well-behaved at each end of $\Sigma$, i.e.  if there exists a
globally-defined smooth function $\hV:\hSigma\rightarrow\R$ such that
$ V=\hV|_\Sigma$.

\subsection{Behavior near the ends}

The cosmological equations of motion in semi-geodesic coordinates
$(r,\theta)$ on an appropriate vicinity of an end $p\in
\hSigma\setminus \Sigma$ reduce to \cite{alpha}:
\beqa
&&\ddot{r}-3\epsilon_p \alpha \left( \frac{C_p}{4\pi }\right) ^{2}e^{2\epsilon_p r}
\dot{\theta}^{2}+3H\dot{r}+\frac{1}{3\alpha }\partial _{r} V =0~~,\\
&&\ddot{\theta}+2\epsilon_p \dot{r}\dot{\theta}+3H\dot{\theta}+\frac{1}{3\alpha }
\left(\frac{4\pi }{C_p}\right) ^{2}e^{-2\epsilon_p r}\partial _{\theta } V
=0~~,
\eeqa
where $C_p$ and $\epsilon_p$ are known constants depending on the type
of end.  Since $\theta$ is periodic, a generic trajectory will spiral
around the ends for any well-behaved scalar potential. Using an
argument similar to that of \cite{Linde3}, we showed in \cite{alpha}
that generalized $\alpha$-attractor models have the same kind of
"universal" behavior as the disk models of \cite{Escher} in the {\em
  naive} one field truncation near each end obtained by
fixing $\theta$. The cosmological behavior away from the ends is 
much more subtle than that of ordinary $\alpha$-attractors;
some of its qualitative features were discussed in \cite{alpha}. 
Various examples are discussed in \cite{elem,modular}.

\section{Examples of trajectories for the  hyperbolic triply punctured sphere}

Consider those generalized $\alpha$-attractor models for which the
scalar manifold $\Sigma$ is the triply-punctured Riemann sphere
(a.k.a. the modular curve) $Y(2)\eqdef\C \P^1\setminus
\{p_1,p_2,p_3\}$. This is diffeomorphic with the doubly-punctured
complex plane endowed with its complete hyperbolic metric $\dd
s^2=\rho(\zeta,\bar{\zeta})^2\dd \zeta^2$, where:
\be
\rho(\zeta,{\bar \zeta})=\frac{\pi}{8|\zeta(1-\zeta)|} \, \frac{1}{\Re [\cK(\zeta)\cK(1-\bar{\zeta})]}~~,~~
\cK(\zeta)=\int_{0}^1 \frac{\dd t}{\sqrt{(1-t^2)(1-\zeta t^2)}}~.
\ee
Each of the three punctures $p_i$ corresponds to a cusp end, so the
end compactification is ${\hSigma}=\C\P^1\simeq\rS^2$.  The
surface $Y(2)$ is uniformized by the principal congruence
subgroup of level $2$, $
\Gamma(2)\!=\! \Big\{A\!=\!\left[\begin{array}{cc}a & b \\ c & d \end{array}\right]\in \PSL(2,\Z)~| ~a,d=\odd\,,~b,c=\mathrm{even}\Big\} \,
$,
with  uniformization map given by the elliptic modular lambda function 
$
\pi_\H(\tau)\!\equiv\!\lambda(\tau)\!=\!\frac{\wp_\tau(\frac{1+\tau}{2})-\wp_\tau(\frac{\tau}{2})}{\wp_\tau(\frac{1}{2})-\wp_\tau(\frac{\tau}{2})}~,
$
where $\wp$ is the Weierstrass elliptic function of modulus $\tau$.  A
fundamental polygon for the action of $\Gamma(2)$ on $\H$ is given by
the hyperbolic quadrilateral $\fD_{\mathbb{H}} =\{\tau \in \H| -1 <
\Re \tau < 0, |\tau+\frac{1}{2}|> \frac{1}{2} \}\cup \{\tau \in \H| 0
\leq \Re \tau < 1, |\tau-\frac{1}{2}|> \frac{1}{2}\}$.  

\

\noindent Consider the following two globally well-behaved scalar potentials:
\beqa
\label{V0}
\hV_0(\psi,\theta)&\eqdef&M_0(1+\sin\psi\cos\theta)~~,\\
\label{Vp}
\hV_+(\psi)&\eqdef&M_0\cos^2\frac{\psi}{2}~~,
\eeqa
where $M_0\eqdef M\sqrt{2/3}$ and $\psi,\theta$ are spherical
coordinates on the end compactification $\hSigma=\rS^2$. 
Fixing $\alpha=\frac{1}{3}$ and choosing the initial conditions $\tau_0$ and
$\tv_0\eqdef\dot\tphi(t_0)$ given in Table \ref{tab:InCond}, we
compute \cite{modular} the lifted trajectories on the Poincar\'e 
half-plane with coordinate $\tau=x+\i y$ and then project them to $Y(2)$
(see Figs. 1 and 2). The potentials $\hV_0$ and respectively $\hV_+$
correspond to $\tV_0$ and $\tV_+$ on $\H$ and to
$V_0$ and $V_+$ on $Y(2)$.

\begin{figure}
\label{fig:Phi0}
\centering \includegraphics[width=11cm]{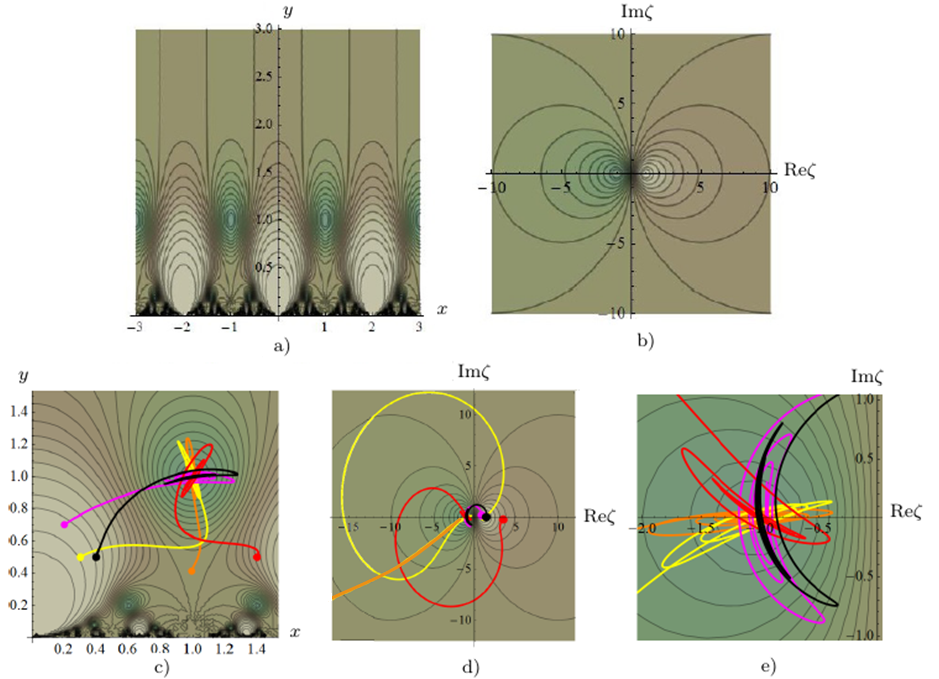}
\caption{a) Level plot of the lifted potential $\tV_0$ on $\H$. b)
  Level plot of $V_0$ on $Y(2)$. c) Lifted trajectories on $\H$, 
  with initial conditions given in Table 1. d)
  Projected trajectories on $Y(2)$, where the orange trajectory is too
  long to fit into the plot at the scale shown. e) Detail of the
  spiral ends of the trajectories on $Y(2)$. The beginning points of the
  trajectories are indicated by fat dots. In all figures, dark green
  indicates minima of the potential while light brown indicates maxima.}
\end{figure}
For the potential $\tV_0$, we find that the red, magenta, yellow and
orange trajectories start in inflationary regime (see Fig. 3), but
computations show they have small number of e-folds (less than 5); on
the other hand, the black trajectory is not inflationary. For
potential $\tV_+$, we find that the red, yellow and orange trajectories (see Fig. 4)
start in inflationary regime, while the magenta and black trajectories
are not inflationary. The orange trajectory has 50 e-folds and using very small
variations of its initial conditions given in Table 1 we can easily
find other trajectories with 50-60 e-folds; this shows that
generalized $\alpha$ attractors with $\Sigma=Y(2)$ can produce
phenomenologically realistic predictions. The number of e-folds is
given by $N=\int_{0}^{t_I}H(t)\dd t$, where $t_I$ is the
inflationary period (the duration of the first inflationary regime).

\begin{table}
\caption{Initial conditions on the Poincar\'e half-plane.}
\label{tab:InCond}       
\begin{tabular}{p{2.4cm}p{2.4cm}p{1.7cm}}
\hline\noalign{\smallskip}
trajectory & $\tau_0$ & $\tv_0$ \\
\noalign{\smallskip}\svhline\noalign{\smallskip}
black  & $0.4+0.5\i$  & $0.3+\i$ \\
red  & $1.4+0.5\i$ & $0.1+0.2\i$  \\
magenta & $0.2+0.7\i$ & $0.7+0.5\i$  \\
yellow & $0.3+0.5\i$  & $0$\\
orange & $0.99+0.415\i$  & $0$\\
\noalign{\smallskip}\hline\noalign{\smallskip}
\end{tabular}
\end{table}

\begin{figure}
\label{fig:PhiPlus}
\centering \includegraphics[width=12cm]{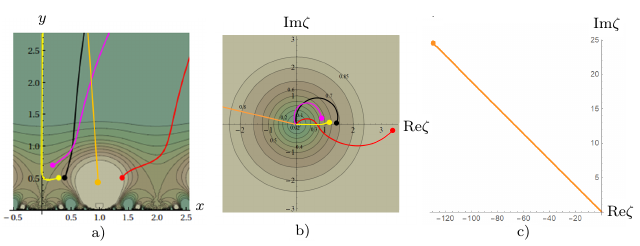}
\caption{a) Level plots of the lifted potential $\tV_+$ on $\H$ and
  the lifted trajectories with initial conditions given in Table 1.
  b) Level plots of $V_+$ on $Y(2)$ and the corresponding projected
  trajectories. c) The full orange trajectory projected on $Y(2)$.}
\end{figure}

\begin{figure}
\label{fig:Htraj0}
	\centering \includegraphics[width=2.9cm]{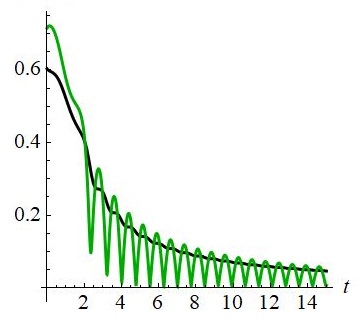}\includegraphics[width=2.9cm]{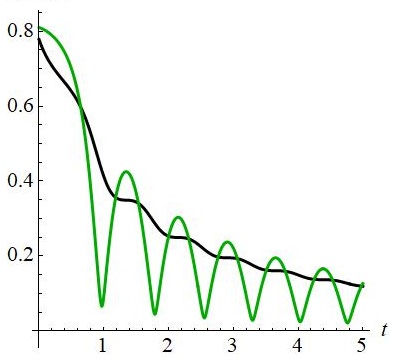}
\includegraphics[width=2.9cm]{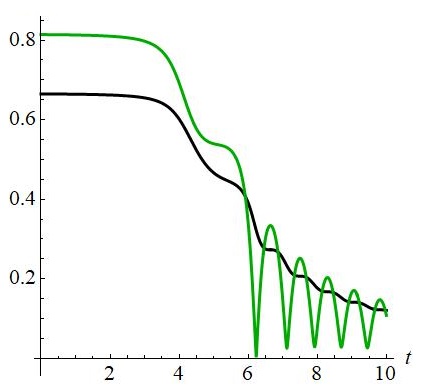}\includegraphics[width=2.9cm]{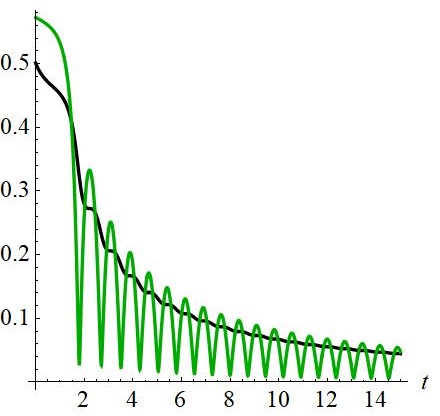}
\caption{Plot of $H(t)/\sqrt{M_0}$ (black) and $H_c(t)/\sqrt{M_0}$ (green) for the
 red, magenta, yellow and orange trajectories for the potential $\tV_0$.}
\end{figure}
\begin{figure}
\label{fig:HtrajPlus}
	\centering \includegraphics[width=3.3cm]{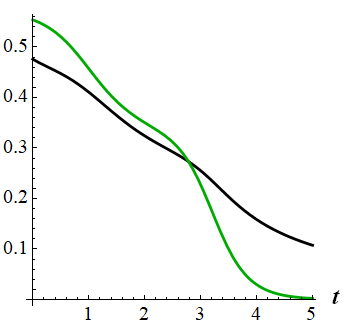}~~\includegraphics[width=3.3cm]{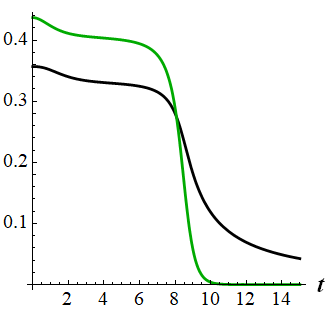}~~
\includegraphics[width=3.3cm]{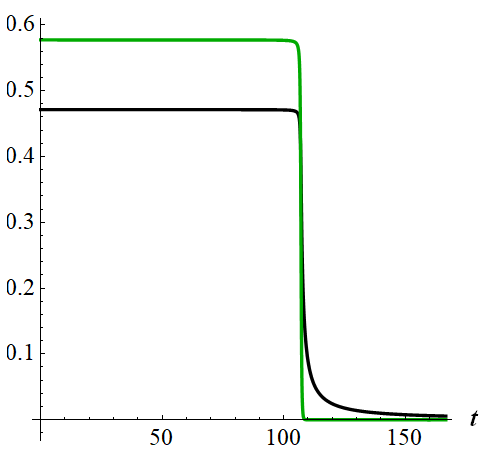}~
\caption{Plot of $H(t)/\sqrt{M_0}$ (black) and $H_c(t)/\sqrt{M_0}$ (green) for the
 red, yellow and orange trajectories in the potential $\tV_+$. 
 The red and yellow trajectorie have small number 
 of e-folds (less than 2), but the orange trajectory has 50 e-folds.}
\end{figure}

\section{Conclusions}

We proposed \cite{alpha, modular, elem} a wide generalization of
two-field $\alpha$-attractor models obtained by promoting the scalar
manifold from the Poincar\'e disk to an arbitrary geometrically finite
non-compact hyperbolic surface and a procedure for studying such
models through uniformization techniques. Our models are parameterized
by a constant $\alpha>0$, by the choice of a surface group $\Gamma$
and of a smooth scalar potential $V$. They have the same universal
behavior as ordinary $\alpha$-attractors in the naive one-field
truncation near each end, provided that the scalar potential is
well-behaved near that end.

\begin{acknowledgement}
This work was supported by grant IBS-R003-S1.
\end{acknowledgement}

\end{document}